\documentclass[10pt,a4paper]{article}

\usepackage[T1]{fontenc}
\usepackage[utf8]{inputenc}
\usepackage[a4paper,margin=1.7cm]{geometry}
\usepackage{microtype}
\usepackage{graphicx}
\usepackage{amsmath,amssymb}
\usepackage{booktabs}
\usepackage{array}
\usepackage{tabularx}
\usepackage{multirow}
\usepackage{float}
\usepackage{changepage}
\usepackage[numbers,sort&compress]{natbib}
\usepackage[hidelinks]{hyperref}
\hypersetup{
  pdftitle={Load-Path Redistribution and Damage Asymmetry in Reinforced Concrete Beams under Eccentric Drop-Weight Impact: A Coupled SPH--FEM Study},
  pdfauthor={Ziqi Gao, Chi Lu, Yoshimi Sonoda, Hiroki Tamai},
  pdfsubject={Coupled SPH--FEM analysis of reinforced concrete beams under eccentric drop-weight impact},
  pdfkeywords={reinforced concrete beam, impact loading, SPH--FEM coupling}
}

\newcolumntype{C}{>{\centering\arraybackslash}X}
\newcolumntype{L}{>{\raggedright\arraybackslash}X}
\newcolumntype{R}{>{\raggedleft\arraybackslash}X}
\newcommand{\tablesize}[1]{#1}
\newlength{\extralength}
\newlength{\fulllength}
\title{\textbf{Load-Path Redistribution and Damage Asymmetry in Reinforced Concrete Beams under Eccentric Drop-Weight Impact: A Coupled SPH--FEM Study}}
\author{Ziqi Gao$^{1}$, Chi Lu$^{1,*}$, Yoshimi Sonoda$^{1}$, and Hiroki Tamai$^{1}$\\[0.5em]
\small $^{1}$Department of Civil Engineering, Kyushu University, 744 Motooka,\\
\small Nishi-ku, Fukuoka 819-0395, Japan\\[0.25em]
\small $^{*}$Corresponding author: \href{mailto:luchi@doc.kyushu-u.ac.jp}{luchi@doc.kyushu-u.ac.jp}}
\date{}

\begin{document}
\maketitle

\begin{center}
\begin{minipage}{0.94\textwidth}
\small
\textbf{Author-posted version.}
This is an author-posted version of the article published in \textit{Applied Sciences}, 2026, 16(13), 6700.
The version of record is available at \href{https://doi.org/10.3390/app16136700}{https://doi.org/10.3390/app16136700}.
The published article is distributed under the Creative Commons Attribution 4.0 International License.
\end{minipage}
\end{center}

\begin{abstract}
Reinforced concrete (RC) beams under impact are commonly assessed using central-impact configurations, but practical impacts may deviate from midspan and create unequal shear spans. This study investigates how impact eccentricity changes force transfer and damage development using a validated coupled smoothed particle hydrodynamics--finite element method (SPH--FEM) model. Concrete is modeled with SPH particles, while reinforcement, supports, and the impactor are modeled with FEM solid elements. After validation against central drop-weight tests, full-span eccentric-impact cases are compared with matched short-span references. The first contact-force peak changes only slightly with eccentricity, whereas the response distribution changes clearly. At the largest eccentricity, shorter-span shear reaches up to 2.23 times the central-impact value, showing shear-dominated redistribution. Absorbed energy per unit length follows the same trend in shorter-span, reaching up to 4.29 times the longer-span-side value. Matched references show that full-span eccentric beams can develop up to \(18.4~\mathrm{kN}\) higher local shear than symmetric short-span beams. Damage fields shift from symmetric central damage to asymmetric shorter-span-side damage with clearer fragmentation in low-strength cases. Eccentric impact should therefore be evaluated as a full-span shear-transfer and damage-asymmetry problem.
\end{abstract}

\noindent\textbf{Keywords:} reinforced concrete beam; impact loading; SPH--FEM coupling

\medskip
\noindent\textbf{Practical application:} The results can support assessment of RC beams subjected to off-midspan impact, particularly where shorter-span-side shear transfer and asymmetric damage need to be evaluated.

\medskip

\section{Introduction}

Concrete and reinforced concrete (RC) members may be subjected to localized impact from falling objects, rockfall, vehicle or equipment collision, debris, and construction activities. Impact response differs from quasi-static response because the load is short in duration, spatially concentrated, and strongly coupled with inertia, stress-wave propagation, local crushing, and strain-rate effects. Early drop-weight studies established experimental procedures for concrete impact and discussed rate-dependent fracture and energy absorption \cite{Bentur1986,Banthia1989}. Material-level studies further showed that concrete in compression and tension, as well as reinforcing steel, changes its mechanical response with loading rate \cite{BischoffPerry1991,MalvarRoss1998,Malvar1998,SoroushianChoi1987}. At the structural scale, studies on missile impact, RC slab impact, and impact-induced damage showed that local contact damage, wave propagation, and spatial damage distribution can control the observed response \cite{Li2005,Delhomme2007,Zineddin2007,Hering2020}.

RC beams under drop-weight or concentrated impact are useful benchmark members because their span, cross-section, reinforcement, support condition, and response measurements can be controlled. Tests on shear-failure-type beams without shear reinforcement showed that local shear damage can become critical under falling-weight impact \cite{KishiMikami2002}. Analytical and experimental studies on RC beams under impact clarified the relation among impact force, displacement response, energy absorption, and damage mode \cite{Fujikake2009,KishiMikami2012}. Other numerical and experimental studies examined the influence of shear reinforcement, reinforcement layout, and nonlinear stress redistribution on impact response \cite{SaatciVecchioShear2009,KishiBhatti2010,Bhatti2009,OzboltSharma2011}. High-rate concentrated-loading studies further showed that the dynamic demand of RC beams cannot be inferred directly from static load-carrying capacity \cite{Cotsovos2008,Cotsovos2010,Adhikary2012}. More recent beam studies have extended this basis to quasi-static and impact loading comparisons, finite element analysis, and residual static or impact capacity after an initial impact event \cite{Soleimani2019,Peterson2022}. These studies make central impact a well-established reference configuration for RC beam impact analysis.

Peak impact force and peak displacement are useful global response measures, but they do not fully describe how impact demand is transferred through a beam. During impact, diagonal cracking, local shear transfer, shear-plug tendency, stress redistribution, and support restraint may develop before a static-type mechanism is formed. Experimental observations by Saatci and Vecchio showed that shear mechanisms can dominate the impact behavior of RC beams \cite{SaatciVecchioShear2009}. Numerical studies have also emphasized the importance of constitutive modelling, shear reinforcement, and nonlinear stress redistribution in reproducing beam impact response \cite{SaatciVecchioNFEM2009,OzboltSharma2011,CotsovosPavlovic2012}. Zhao et al. interpreted beam impact through local and member-level response phases and evaluated dynamic moment and shear demand using a modified-compression-field-theory-based failure criterion \cite{Zhao2016}. This focus on shear demand is consistent with established shear theories based on compression-field concepts, critical shear cracks, and empirical shear-strength evaluation \cite{VecchioCollins1986,Bentz2006,Muttoni2008,Collins2008,Zsutty1968}.

Most available beam-impact benchmarks place the impact point near midspan. This configuration gives a symmetric response in a simply supported beam and is therefore valuable for validation. Practical impact locations, however, may deviate from midspan. Once the impact point moves away from midspan, one shear span becomes shorter and the other becomes longer. The shorter-span side then provides a more direct route for force transfer to the closer support, while the contact section may have a lower bending moment than in central impact. The response may therefore shift from a symmetric two-sided transfer pattern to a shorter-span-side shear-transfer pattern. Studies on localized impact and punching of slabs show that concentrated force transfer can govern local damage away from a global flexural mechanism \cite{Micallef2014,HrynykVecchio2014,Delhomme2007}. Slab impact and damage-quantification studies also show that impact damage is often spatially nonuniform and needs to be described beyond a single global response value \cite{Zineddin2007,Hering2020}. Studies on support-region and inclined-section behavior further indicate that force transfer near supports can become critical when transverse force and bending moment act together \cite{Kos2022}. These findings motivate an eccentric-impact study that links contact-force history with sectional demand, energy localization, and damage asymmetry.

Eccentric drop-weight impact of RC beams is examined using a coupled SPH--FEM model. SPH and related meshfree Lagrangian methods are suitable for impact problems involving large deformation, cracking, fragmentation, and material separation \cite{Johnson1996,RabczukEibl2003,RabczukEiblStempniewski2004}. In the present model, concrete is represented by SPH particles, while reinforcement, supports, and the impactor are represented by FEM solid elements. The model is first checked against central-impact benchmark tests and is then used in an eccentricity-centered simulation program with impact-velocity and concrete-strength branches. The response is evaluated through contact-force pulse response, sectional shear and bending-moment demand, span-normalized absorbed-energy density, matched short-span reference cases, and damage morphology. The analysis examines how off-midspan impact redirects demand to the shorter-span side and whether the redistribution is shear-dominated. The full-span eccentric beam is also compared with a symmetric short-span reference whose half span matches the shorter span of the eccentric beam. This comparison separates local-span effects from full-span interaction.

\section{Numerical Model Validation}

\subsection{Validation benchmark and impact-test configuration}

\begin{figure}[H]
  \centering
  \includegraphics[width=\linewidth]{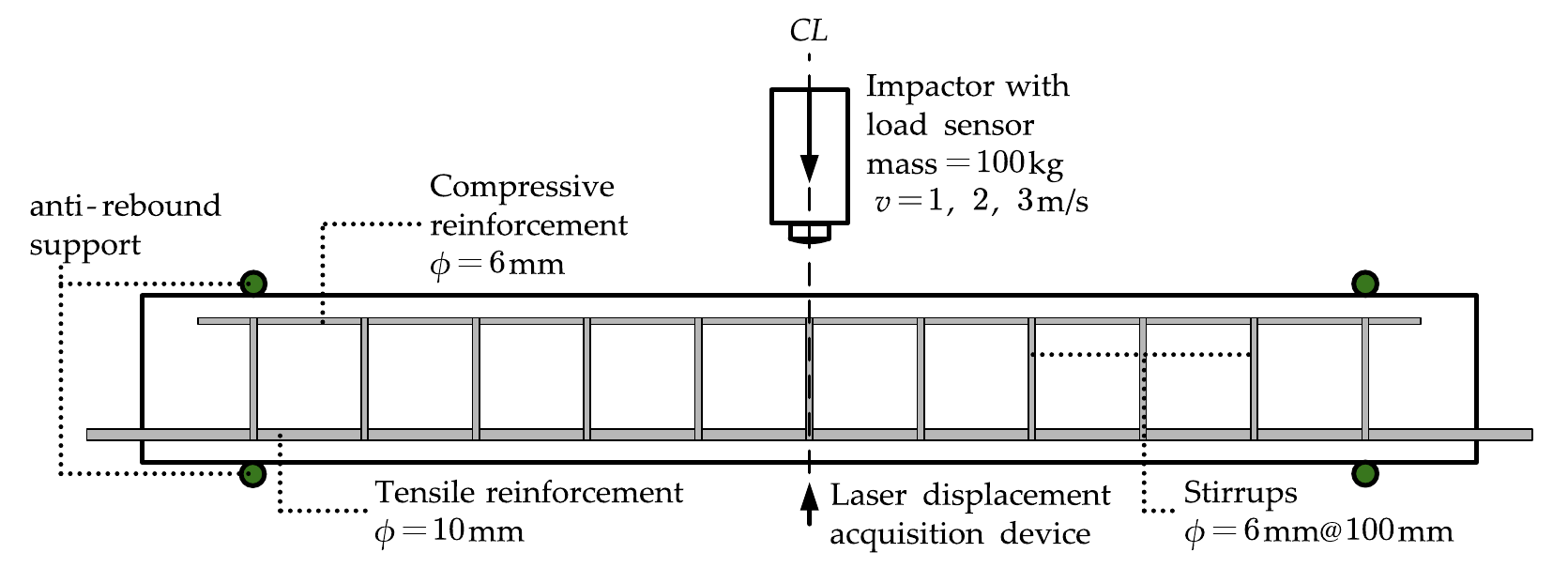}
  \caption{Benchmark beam, reinforcement layout, and impact-test configuration used for model validation.\label{fig:beam_validation_2d}}
\end{figure}

The numerical model was checked against the central drop-weight impact beam tests of Tamai et al.~\cite{Tamai2020ImpactBeamTests}. Figure~\ref{fig:beam_validation_2d} shows the benchmark specimen, reinforcement layout, support arrangement, impactor, and measurement configuration used for validation. The specimen was a simply supported RC beam with dimensions of \(1200~\mathrm{mm} \times 100~\mathrm{mm} \times 150~\mathrm{mm}\) and a support span of \(1000~\mathrm{mm}\). The reference concrete had a compressive strength of \(45~\mathrm{MPa}\) and an elastic modulus of \(30.3~\mathrm{GPa}\). The tensile reinforcement consisted of \(\phi 10~\mathrm{mm}\) bars, while the compression reinforcement and stirrups had a diameter of \(\phi 6~\mathrm{mm}\). The stirrups were spaced at \(100~\mathrm{mm}\).

A \(100~\mathrm{kg}\) drop weight equipped with a load sensor struck the beam at midspan. Three impact velocities, \(1\), \(2\), and \(3~\mathrm{m/s}\), were used for the validation cases. The midspan displacement was measured by a laser displacement device, and anti-rebound fixtures were installed near the support regions. The validation model reproduced the same beam geometry, support span, reinforcement layout, concrete reference properties, impactor mass, loading position, support arrangement, and displacement-measurement point. The numerical material parameters, discretization, interface setting, and explicit time increment are summarized in Table~\ref{tab:model_parameters}.

\subsection{Hybrid SPH--FEM analysis model}

The validation and eccentric-impact simulations were performed using an in-house GPU-parallel explicit SPH--FEM solver implemented with Taichi~\cite{Hu2019Taichi}. The solver stores particle states, FEM nodal and element states, active interface interactions, and material history variables in GPU fields updated by parallel kernels. Concrete was discretized by SPH particles to represent localized crushing, damage spreading, and material separation. Reinforcement, the impactor, and support fixtures were discretized by eight-node solid finite elements, which provide a mesh-based description of the steel skeleton and boundary components.

The SPH and FEM domains were coupled through local interface force exchange. Contact interaction transfers normal compression and tangential sliding resistance between concrete particles and adjacent FEM surfaces. Bond transfer is applied at the reinforcement--concrete interface. The same contact and bond treatment was used in the validation and eccentric-impact simulations.

For a concrete particle \(i\), the semi-discrete equation of motion is written as
\begin{equation}
  m_i \mathbf{a}_i =
  \mathbf{f}^{\mathrm{SPH}}_i+
  \mathbf{f}^{\Gamma}_i ,
  \label{eq:sph_particle_balance}
\end{equation}
where \(m_i\) and \(\mathbf{a}_i\) are the particle mass and acceleration, \(\mathbf{f}^{\mathrm{SPH}}_i\) is the internal force contribution from the SPH concrete domain, and \(\mathbf{f}^{\Gamma}_i\) is the total force transferred from adjacent FEM surfaces. For an FEM node \(a\), the nodal balance is
\begin{equation}
  m_a \ddot{\mathbf{u}}_a =
  \mathbf{f}^{\mathrm{ext}}_a
  -
  \mathbf{f}^{\mathrm{int}}_a
  +
  \mathbf{r}^{\Gamma}_a ,
  \label{eq:fem_nodal_balance}
\end{equation}
where \(m_a\) is the lumped nodal mass, \(\mathbf{f}^{\mathrm{ext}}_a\) and \(\mathbf{f}^{\mathrm{int}}_a\) are the applied and element-resisting nodal forces, and \(\mathbf{r}^{\Gamma}_a\) is the interface reaction assembled from interacting concrete particles. The interface transfer terms are assembled locally with equal and opposite forces. The SPH particles and FEM nodes are then advanced using a common explicit time increment based on their lumped masses.

\begin{figure}[H]
  \begin{adjustwidth}{-\extralength}{0cm}
    \centering
    \includegraphics[width=\fulllength]{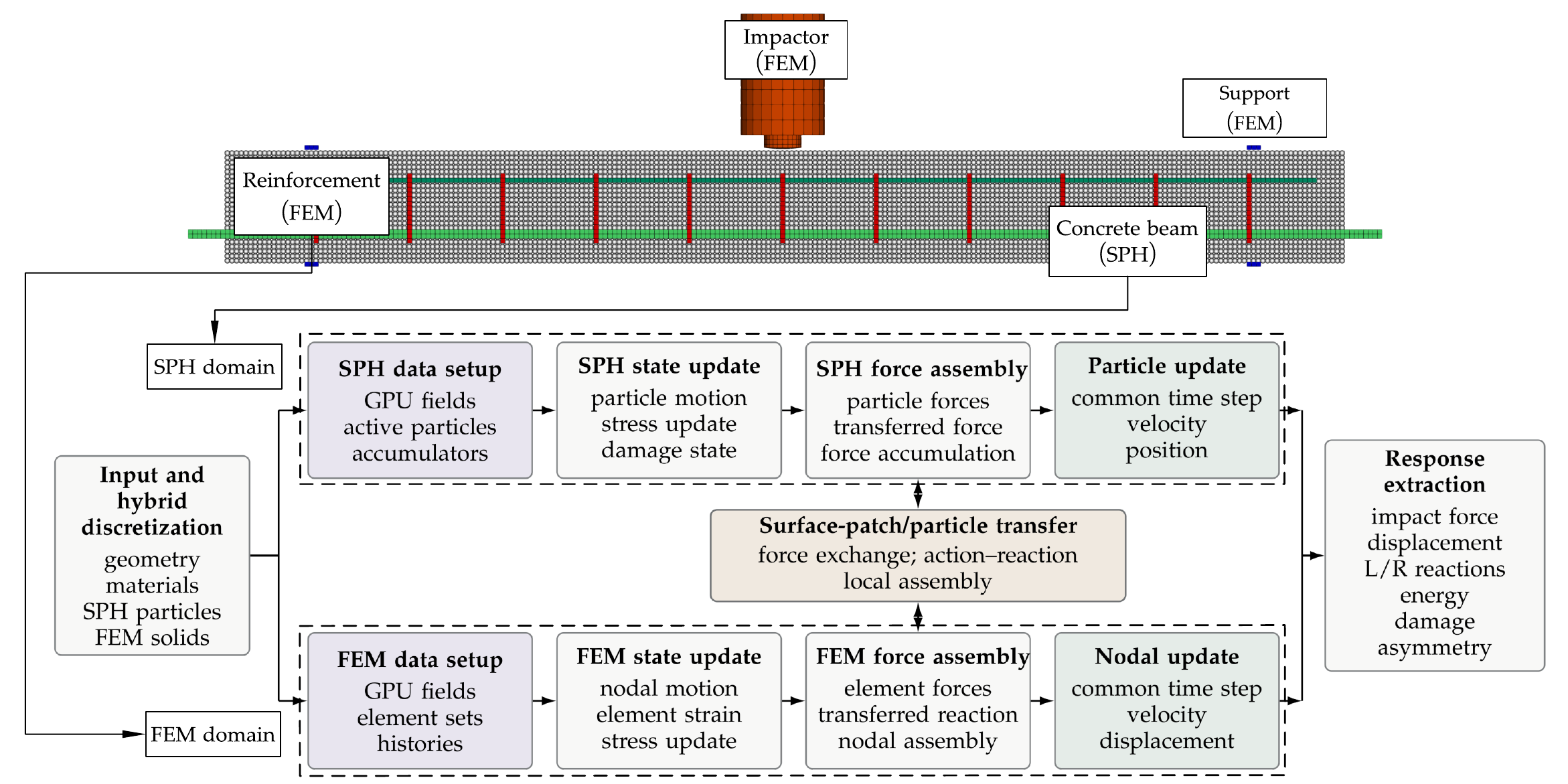}
    \caption{Hybrid SPH--FEM numerical model and GPU-parallel computational workflow used for impact simulations. The upper part shows the SPH concrete beam and the FEM reinforcement, impactor, and supports. The lower part shows Taichi data preparation, SPH and FEM state updates, interface force transfer, explicit particle and nodal updates, and response quantities extracted for eccentric-impact analysis.\label{fig:sph_fem_workflow}}
  \end{adjustwidth}
\end{figure}

Figure~\ref{fig:sph_fem_workflow} summarizes the hybrid numerical model and the execution sequence of the GPU-parallel solver. The upper part identifies the SPH concrete beam and the FEM reinforcement, impactor, and supports used in the validation model. The lower part shows the calculation sequence from Taichi data preparation to SPH and FEM state updates, local force assembly, interface force transfer, explicit particle and nodal updates, and response extraction. The extracted histories and field quantities are later used to evaluate contact-force pulse response, sectional demand, energy distribution, and damage asymmetry.

\subsection{Material models and numerical parameters}

Concrete was described using the continuous surface cap model (CSCM)~\cite{Murray2007CSCM}. The model accounts for pressure-dependent response, tensile and shear damage, compressive compaction, and fracture-energy-based softening regularization. For the validation benchmark, the CSCM input parameters were generated from the measured compressive strength and elastic modulus of the specimen concrete. In the concrete-strength branch, the prescribed compressive strength was changed, and the derived CSCM parameters were updated consistently.

Reinforcement was modeled using a von Mises \(J_2\) elastoplastic model. The impactor and support fixtures were represented by solid FEM bodies. The impactor density was assigned as an equivalent value so that the FEM body reproduced the total drop-weight mass. Steel--concrete bond was represented by a tabulated envelope with unloading and reloading rules. The same contact and bond settings were used for both the validation cases and the eccentric-impact simulations.

Table~\ref{tab:model_parameters} summarizes the main geometry, material, discretization, interface, and time-integration parameters of the baseline model. The same baseline discretization and interface settings were used for the central-impact validation cases and for the eccentric-impact simulation program. The variables changed in the parametric study were the prescribed impact velocity, concrete compressive strength, and impact eccentricity, as defined in Section~3.

\begin{table}[H]
  \caption{Geometry, material, and numerical parameters of the baseline model.\label{tab:model_parameters}}
  \centering
  \tablesize{\footnotesize}
  \begin{tabularx}{\textwidth}{LLL}
    \toprule
    \textbf{Item} & \textbf{Value} & \textbf{Note} \\
    \midrule
    Beam geometry & \(1200~\mathrm{mm} \times 100~\mathrm{mm} \times 150~\mathrm{mm}\), support span \(1000~\mathrm{mm}\) & Benchmark specimen \\
    Reinforcement layout & Tensile bars \(\phi 10~\mathrm{mm}\), compression bars and stirrups \(\phi 6~\mathrm{mm}\) & Stirrups spaced at \(100~\mathrm{mm}\) \\
    Impactor & \(100~\mathrm{kg}\) drop weight & Equivalent FEM density used to reproduce mass \\
    Concrete, reference case & CSCM, \(\rho=2350~\mathrm{kg/m^3}\), \(E=30.3~\mathrm{GPa}\), \(\nu=0.23\), \(f_c=45~\mathrm{MPa}\) & Strength varied in strength branch \\
    Reinforcement & \(J_2\) elastoplasticity, \(\rho=7800~\mathrm{kg/m^3}\), \(E=188~\mathrm{GPa}\), \(\nu=0.30\), \(f_y=345~\mathrm{MPa}\) & Longitudinal bars and stirrups \\
    Impactor and supports & Solid FEM bodies, \(E=210~\mathrm{GPa}\) for the impactor, \(E=188~\mathrm{GPa}\) for supports, \(\nu=0.30\) & Boundary and loading components \\
    Discretization & Concrete SPH particles, \(5.0~\mathrm{mm}\); reinforcement FEM Hex8, \(5.0~\mathrm{mm}\); support FEM Hex8, \(5.0~\mathrm{mm}\); impactor FEM Hex8, \(6.5~\mathrm{mm}\) average & Baseline mesh \\
    Model size & \(112824\) concrete particles, \(3032\) reinforcement elements, \(240\) support elements, \(1236\) impactor elements & Validation model \\
    Interface setting & Penalty contact and tabulated steel--concrete bond envelope & Same setting for all cases \\
    Explicit time increment & \(1.0 \times 10^{-7}~\mathrm{s}\) & Common time step \\
    \bottomrule
  \end{tabularx}
  \tablesize{}
\end{table}

\subsection{Validation results and response-history comparison}

\begin{figure}[H]
  \centering
  \includegraphics[width=\linewidth]{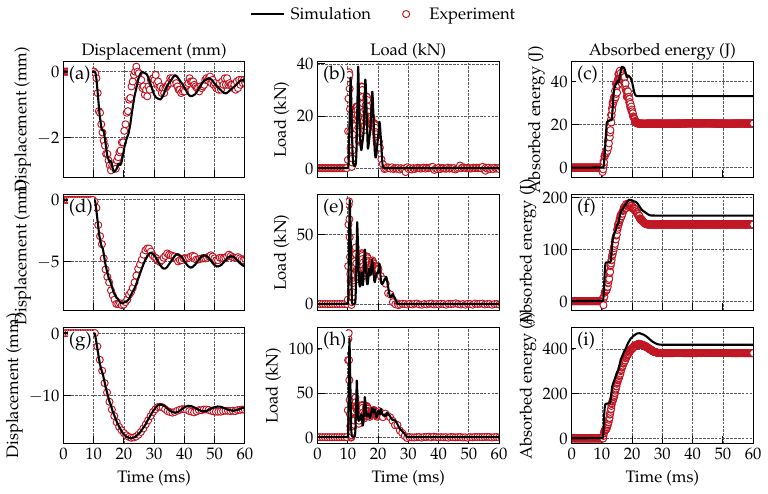}
  \caption{Validation of displacement, impact-force, and absorbed-energy histories for representative benchmark cases. Panels (a--c), (d--f), and (g--i) correspond to I1-N, I2-N, and I3-N. The three columns show displacement, impact force, and absorbed energy. Red open circles denote experimental results, and black lines denote simulations.\label{fig:validation_histories}}
\end{figure}

\begin{figure}[H]
  \centering
  \includegraphics[width=0.80\linewidth]{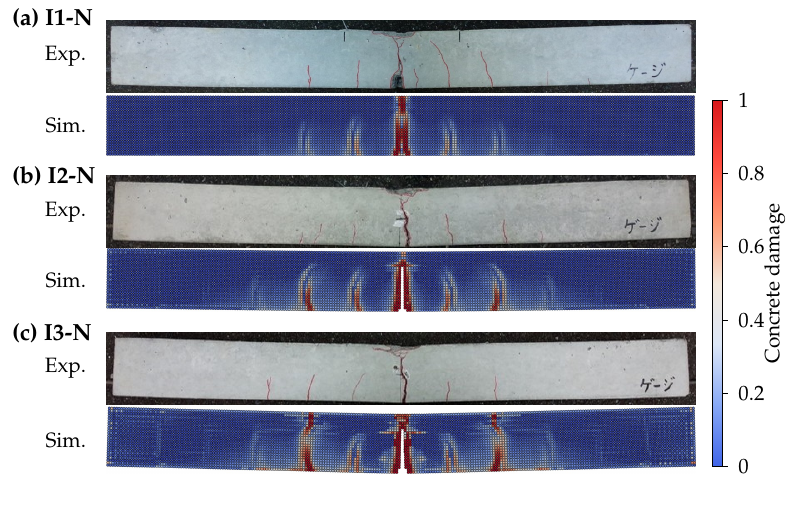}
  \caption{Experimental crack patterns and simulated concrete damage fields for representative benchmark cases: (a) I1-N, (b) I2-N, and (c) I3-N. In each panel, the upper image shows the post-impact experimental crack pattern and the lower image shows the simulated concrete damage field. The color bar denotes the concrete damage variable.\label{fig:validation_damage_comparison}}
\end{figure}

The representative benchmark cases I1-N, I2-N, and I3-N were used to validate the impact response of the coupled SPH--FEM model. The validation is reported in three levels. Table~\ref{tab:validation_summary} compares the main scalar response quantities, Figure~\ref{fig:validation_histories} compares the displacement, impact-force, and absorbed-energy histories, and Figure~\ref{fig:validation_damage_comparison} compares the observed crack patterns with the simulated damage fields. In Table~\ref{tab:validation_summary}, \(v\) denotes the impact velocity, \(u_{\max}\) denotes the maximum displacement, \(F_1\) denotes the first peak impact force, and \(T_p\) denotes the load-pulse duration. The error is reported as the absolute relative difference, \(|\mathrm{sim}-\mathrm{exp}|/\mathrm{exp}\times100\%\).

\begin{table}[H]
  \caption{Validation summary for representative benchmark cases.\label{tab:validation_summary}}
  \begin{adjustwidth}{-\extralength}{0cm}
    \tablesize{\footnotesize}
    \begin{tabularx}{\fulllength}{LCCCCCCCCCC}
      \toprule
      \textbf{Case} & \boldmath{\(v\)} & \boldmath{\(u_{\max}^{\mathrm{exp}}\)} & \boldmath{\(u_{\max}^{\mathrm{sim}}\)} & \textbf{Err.} & \boldmath{\(F_1^{\mathrm{exp}}\)} & \boldmath{\(F_1^{\mathrm{sim}}\)} & \textbf{Err.} & \boldmath{\(T_p^{\mathrm{exp}}\)} & \boldmath{\(T_p^{\mathrm{sim}}\)} & \textbf{Err.} \\
      & \textbf{(m/s)} & \textbf{(mm)} & \textbf{(mm)} & \textbf{(\%)} & \textbf{(kN)} & \textbf{(kN)} & \textbf{(\%)} & \textbf{(ms)} & \textbf{(ms)} & \textbf{(\%)} \\
      \midrule
      I1-N & \(1\) & \(3.05\) & \(3.05\) & \(0.1\) & \(36.8\) & \(34.7\) & \(5.7\) & \(12.0\) & \(11.5\) & \(4.1\) \\
      I2-N & \(2\) & \(8.60\) & \(8.38\) & \(2.6\) & \(74.0\) & \(75.1\) & \(1.5\) & \(15.2\) & \(16.2\) & \(6.7\) \\
      I3-N & \(3\) & \(16.95\) & \(16.96\) & \(0.1\) & \(118.0\) & \(112.1\) & \(4.9\) & \(18.9\) & \(19.5\) & \(3.0\) \\
      \bottomrule
    \end{tabularx}
    {\raggedright\footnotesize\noindent\textit{Note:} Superscripts \(\mathrm{exp}\) and \(\mathrm{sim}\) denote experimental and simulated values, respectively. \(T_p\) is measured over the valid post-impact force interval above \(0.5~\mathrm{kN}\) within \(0\)--\(60~\mathrm{ms}\).\par}
    \tablesize{}
  \end{adjustwidth}
\end{table}

Table~\ref{tab:validation_summary} shows good agreement between the experimental and simulated scalar responses. The errors are \(0.1\)--\(2.6\%\) for \(u_{\max}\), \(1.5\)--\(5.7\%\) for \(F_1\), and \(3.0\)--\(6.7\%\) for \(T_p\). These comparisons indicate that the model captures the main displacement magnitude, initial impact-force scale, and force-pulse duration of the benchmark tests. They also establish the central-impact reference before the impact location is shifted in the eccentric-impact simulation program.

For the time-history comparison in Figure~\ref{fig:validation_histories}, absorbed energy is evaluated from the contact work of the impact force over the impact-point displacement. This gives a global force--displacement work history for validation against the measured impact response:
\begin{equation}
  E_{\mathrm{abs}}(t)
  =
  \int_{t_0}^{t} F_c(\tau)\dot{u}(\tau)\,\mathrm{d}\tau
  =
  \int_{u(t_0)}^{u(t)} F_c\,\mathrm{d}u ,
  \label{eq:absorbed_energy_continuous}
\end{equation}
and the cumulative value for sampled histories is evaluated by the trapezoidal form:
\begin{equation}
  E_{\mathrm{abs},k}
  =
  \sum_{j=1}^{k}
  \frac{F_{c,j}+F_{c,j-1}}{2}
  \left(u_j-u_{j-1}\right).
  \label{eq:absorbed_energy_discrete}
\end{equation}
Here, \(E_{\mathrm{abs}}\) is the cumulative absorbed energy, \(t_0\) is the impact onset time, \(t\) is the current time, \(\tau\) is the integration variable, \(F_c\) is the contact force, \(u\) is the impact-point displacement in the loading direction, the overdot denotes the time derivative, and \(j\) and \(k\) denote the time-sample index and the current sample, respectively. This global contact-work history is used only for validation of the measured force--displacement response. The regional absorbed-energy quantities used later for eccentric-impact energy-density analysis are extracted separately from the field histories.

Figure~\ref{fig:validation_histories} compares the displacement, impact-force, and absorbed-energy histories. The simulation reproduces the displacement histories and the timing and magnitude of the main impact-force pulses for the three benchmark velocities. Short-duration force oscillations differ locally because contact-force histories are sensitive to local crushing, contact stiffness, and signal processing. For I1-N, the absorbed-energy difference after the first energy peak is associated with the rebound stage. The simulated rebound occurs approximately \(2~\mathrm{ms}\) later than the measured rebound, which affects the small residual energy in the lowest-velocity case. At higher velocities, inertia, local crushing, and concrete damage occupy a larger part of the response, and the absorbed-energy histories become closer to the experimental results. The agreement in displacement, first peak force, load-pulse duration, and absorbed-energy evolution provides the validation basis for the eccentric-impact simulations conducted in this study.

Figure~\ref{fig:validation_damage_comparison} compares the experimental crack patterns with the simulated concrete damage fields. The experimental photographs show dominant cracking near the impact section together with secondary flexural cracks along the span. The simulated damage fields reproduce the main central damage zone and show a wider damage distribution as the impact velocity rises from I1-N to I3-N. The comparison is interpreted qualitatively because the photographs record visible surface cracks, whereas the numerical contours represent a continuous concrete damage variable. The model therefore captures the damage features needed for the eccentric-impact simulations.

\section{Eccentric-Impact Simulation Program}

\subsection{Impact eccentricity and span geometry}

The support span is \(L=x_R-x_L\), where \(x_L\) and \(x_R\) are the left and right support coordinates. The midspan coordinate is
\[
  x_{\mathrm{mid}}=\frac{x_L+x_R}{2}.
\]
The impact eccentricity is the distance from the impact point to midspan:
\begin{equation}
  e = |x_i-x_{\mathrm{mid}}| .
  \label{eq:eccentricity_abs}
\end{equation}

The distances from the impact point to the left and right supports are
\begin{equation}
  a_L=x_i-x_L,
  \qquad
  a_R=x_R-x_i .
  \label{eq:left_right_spans}
\end{equation}
The shorter and longer impact-to-support distances are recorded as
\begin{equation}
  a_n=\min(a_L,a_R),
  \qquad
  a_f=\max(a_L,a_R).
  \label{eq:near_far_spans}
\end{equation}
In the discussion below, the two sides are described directly as the shorter-span side and the longer-span side.

For the full-span beam, \(L=1000~\mathrm{mm}\). The central-impact case has \(e=0\) and equal distances of \(500~\mathrm{mm}\) from the impact point to the two supports. The eccentricity levels \(\mathrm{E1}\), \(\mathrm{E2}\), and \(\mathrm{E3}\) correspond to \(e=100\), \(200\), and \(300~\mathrm{mm}\), respectively. If the impact point is shifted to the right of midspan, the distances from the impact point to the left and right supports are \((a_L,a_R)=(600,400)\), \((700,300)\), and \((800,200)~\mathrm{mm}\). The corresponding shorter-span distances are \(a_n=400\), \(300\), and \(200~\mathrm{mm}\), and the longer-span distances are \(a_f=600\), \(700\), and \(800~\mathrm{mm}\). A shift to the left side gives the mirrored geometry with the same \(e\), \(a_n\), and \(a_f\) values.

\begin{figure}[H]
  \centering
  \includegraphics[width=\linewidth]{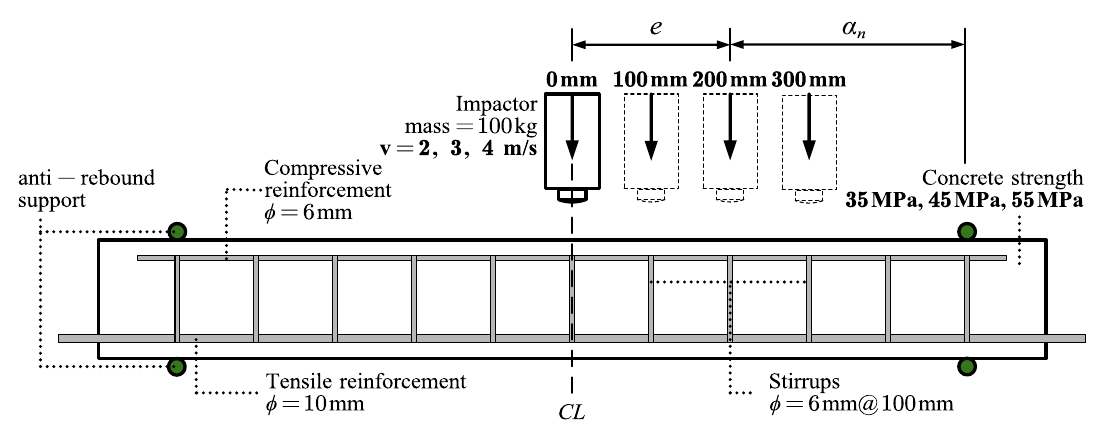}
  \caption{Eccentricity levels and simulation parameters for the full-span eccentric-impact cases.\label{fig:eccentricity_regions}}
\end{figure}

Figure~\ref{fig:eccentricity_regions} shows the full-span beam model, the eccentricity levels, the velocity branch, and the concrete-strength branch used in the eccentric-impact simulations.

\subsection{Full-span case nomenclature and parameter matrix}

\begin{table}[H]
  \caption{Parameter matrix of processed full-span simulation cases grouped by parameter group.\label{tab:simulation_matrix}}
  \centering
  \tablesize{\footnotesize}
  \begin{tabularx}{\textwidth}{LLCCCCC}
    \toprule
    \textbf{Parameter group} & \textbf{Case} & \boldmath{\(v\)} & \boldmath{\(f_c\)} & \boldmath{\(e\)} & \boldmath{\(a_n\)} & \boldmath{\(a_f\)} \\
    & & \textbf{(m/s)} & \textbf{(MPa)} & \textbf{(mm)} & \textbf{(mm)} & \textbf{(mm)} \\
    \midrule
    \multirow{12}{*}{Velocity} & \(\mathrm{S2C45E0}\) & \(2\) & \(45\) & \(0\) & \(500\) & \(500\) \\
    & \(\mathrm{S2C45E1}\) & \(2\) & \(45\) & \(100\) & \(400\) & \(600\) \\
    & \(\mathrm{S2C45E2}\) & \(2\) & \(45\) & \(200\) & \(300\) & \(700\) \\
    & \(\mathrm{S2C45E3}\) & \(2\) & \(45\) & \(300\) & \(200\) & \(800\) \\
    & \(\mathrm{S3C45E0}\) & \(3\) & \(45\) & \(0\) & \(500\) & \(500\) \\
    & \(\mathrm{S3C45E1}\) & \(3\) & \(45\) & \(100\) & \(400\) & \(600\) \\
    & \(\mathrm{S3C45E2}\) & \(3\) & \(45\) & \(200\) & \(300\) & \(700\) \\
    & \(\mathrm{S3C45E3}\) & \(3\) & \(45\) & \(300\) & \(200\) & \(800\) \\
    & \(\mathrm{S4C45E0}\) & \(4\) & \(45\) & \(0\) & \(500\) & \(500\) \\
    & \(\mathrm{S4C45E1}\) & \(4\) & \(45\) & \(100\) & \(400\) & \(600\) \\
    & \(\mathrm{S4C45E2}\) & \(4\) & \(45\) & \(200\) & \(300\) & \(700\) \\
    & \(\mathrm{S4C45E3}\) & \(4\) & \(45\) & \(300\) & \(200\) & \(800\) \\
    \multirow{8}{*}{Strength} & \(\mathrm{S4C35E0}\) & \(4\) & \(35\) & \(0\) & \(500\) & \(500\) \\
    & \(\mathrm{S4C35E1}\) & \(4\) & \(35\) & \(100\) & \(400\) & \(600\) \\
    & \(\mathrm{S4C35E2}\) & \(4\) & \(35\) & \(200\) & \(300\) & \(700\) \\
    & \(\mathrm{S4C35E3}\) & \(4\) & \(35\) & \(300\) & \(200\) & \(800\) \\
    & \(\mathrm{S4C55E0}\) & \(4\) & \(55\) & \(0\) & \(500\) & \(500\) \\
    & \(\mathrm{S4C55E1}\) & \(4\) & \(55\) & \(100\) & \(400\) & \(600\) \\
    & \(\mathrm{S4C55E2}\) & \(4\) & \(55\) & \(200\) & \(300\) & \(700\) \\
    & \(\mathrm{S4C55E3}\) & \(4\) & \(55\) & \(300\) & \(200\) & \(800\) \\
    \bottomrule
  \end{tabularx}
  \tablesize{}
\end{table}

The full-span eccentric-impact simulations were arranged to examine how impact velocity, concrete compressive strength, and impact eccentricity change the response of the validated RC beam model. The case names follow the format \(\mathrm{S}v\mathrm{C}f\mathrm{E}k\). Here, \(\mathrm{S}v\) gives the impact velocity in \(\mathrm{m/s}\), \(\mathrm{C}f\) gives the concrete compressive strength in \(\mathrm{MPa}\), and \(\mathrm{E}k\) gives the eccentricity level. For example, \(\mathrm{S4C35E1}\) denotes a full-span beam impacted at \(v=4~\mathrm{m/s}\), with \(f_c=35~\mathrm{MPa}\) and \(e=100~\mathrm{mm}\). The levels \(\mathrm{E0}\), \(\mathrm{E1}\), \(\mathrm{E2}\), and \(\mathrm{E3}\) correspond to \(e=0\), \(100\), \(200\), and \(300~\mathrm{mm}\), respectively.

Table~\ref{tab:simulation_matrix} lists the processed full-span simulation cases. The velocity group changes the impact velocity from \(2\) to \(4~\mathrm{m/s}\) while keeping the concrete compressive strength at \(45~\mathrm{MPa}\). The strength group changes the concrete compressive strength to \(35\) and \(55~\mathrm{MPa}\) at the fixed impact velocity of \(4~\mathrm{m/s}\). The \(\mathrm{S4C45E0}\) to \(\mathrm{S4C45E3}\) cases belong to the velocity group and also provide the \(45~\mathrm{MPa}\) reference level for the strength comparison, so they are listed only once.

\subsection{Response indicators and data extraction}

The eccentric-impact response is evaluated from four types of output: contact-force histories, sectional shear and bending-moment histories, absorbed energy on the two sides of the impact point, and damage-field snapshots. Together, these outputs show whether the shorter-span side develops larger sectional demand, absorbs more energy per unit length, and shows more visible damage than the longer-span side. This subsection defines how these outputs are extracted and converted into the response quantities used in the full-span discussion.

For contact-force histories, \(F_1\) is the first force peak after impact onset. For the eccentric-impact cases, \(T_p\) is the duration for which the impactor--beam contact force remains present during the main impact event. It is measured from impact onset to contact loss using the same force-history processing rule for all full-span cases.

Sectional force resultants are extracted every \(50~\mathrm{mm}\) along the beam. At each calculation section, the recorded section resultants are the shear force \(V_z(x,t)\) and the bending moment \(M_y(x,t)\). The first \(5~\mathrm{ms}\) after impact are excluded from the envelope extraction to avoid letting the earliest contact shock dominate the sectional comparison. For every \(50~\mathrm{mm}\) section on each side of the impact point, the largest absolute shear or bending-moment value reached after \(t_s=5~\mathrm{ms}\) is recorded. Connecting these maximum values along the beam gives the shear and bending-moment envelopes. The envelopes therefore show the maximum sectional demand that may develop at each position, not the distribution at one single time.

In the formulas, subscript \(n\) refers to the shorter-span side and subscript \(f\) refers to the longer-span side. The peak shear envelopes on the two sides are
\begin{equation}
  V_{n,\max}^{>5\mathrm{ms}}
  =
  \max_{\substack{t\ge t_s\\x~\text{on the shorter-span side}}}
  |V_z(x,t)| ,
  \label{eq:near_shear_peak}
\end{equation}

\begin{equation}
  V_{f,\max}^{>5\mathrm{ms}}
  =
  \max_{\substack{t\ge t_s\\x~\text{on the longer-span side}}}
  |V_z(x,t)| .
  \label{eq:far_shear_peak}
\end{equation}
The corresponding bending-moment envelopes are
\begin{equation}
  M_{n,\max}^{>5\mathrm{ms}}
  =
  \max_{\substack{t\ge t_s\\x~\text{on the shorter-span side}}}
  |M_y(x,t)| ,
  \label{eq:near_moment_peak}
\end{equation}

\begin{equation}
  M_{f,\max}^{>5\mathrm{ms}}
  =
  \max_{\substack{t\ge t_s\\x~\text{on the longer-span side}}}
  |M_y(x,t)| .
  \label{eq:far_moment_peak}
\end{equation}
In the following text and tables, \(V_n\), \(V_f\), \(M_n\), and \(M_f\) are used as shortened names for these four envelope values. For \(\mathrm{E0}\), both halves are \(500~\mathrm{mm}\). When an eccentric case is compared with \(\mathrm{E0}\), the half span on the same side as the eccentric shift is used as the baseline.

Two shear comparisons are made from these envelope values. The ratio \(R_V\) tells whether the shorter-span side carries more shear than the longer-span side in the same case. The amplification \(A_{V,n}\) shows how many times the shorter-span-side shear increases after the impact point moves away from midspan, using the central-impact case with the same velocity and concrete strength as the baseline:
\begin{equation}
  R_V=\frac{V_n}{V_f},
  \qquad
  A_{V,n}(e,v,f_c)=
  \frac{V_n(e,v,f_c)}
  {V_n(0,v,f_c)} .
  \label{eq:main_shear_indicators}
\end{equation}
The bending-moment ratio is calculated in the same way, using the moment envelope on the shorter-span side divided by that on the longer-span side:
\begin{equation}
  R_M=\frac{M_n}{M_f}.
  \label{eq:moment_ratio}
\end{equation}
The ratio indicates whether bending demand follows the same side-wise trend as shear demand.

Absorbed energy is converted to energy per unit span length before comparing the two sides. This is needed because one side becomes shorter and the other becomes longer after the impact point moves away from midspan. The regional energies \(E_n\) and \(E_f\) are extracted from the simulation field-output histories using the section-energy post-processing. Each SPH particle or FEM cell is assigned to one side according to its initial position, and this assignment is kept unchanged during the post-processing. For each output interval, the positive stress-work increment on side \(r\) is accumulated as
\begin{equation}
  \Delta E_{r,m}^{+}
  =
  \sum_{q:\,q~\text{on side }r}
  \max\left[
    V_q\,
    \bar{\boldsymbol{\sigma}}_{q,m}
    :
    \Delta\boldsymbol{\varepsilon}_{q,m},
    0
    \right],
  \qquad
  E_r=\sum_m \Delta E_{r,m}^{+},
  \qquad
  r\in\{n,f\}.
  \label{eq:regional_positive_stress_work}
\end{equation}
Here, \(q\) denotes a concrete SPH particle or reinforcement FEM cell, \(V_q\) is its initial volume, \(\bar{\boldsymbol{\sigma}}_{q,m}\) is the average stress over output interval \(m\), and \(\Delta\boldsymbol{\varepsilon}_{q,m}\) is the strain increment over the same interval. Only positive increments are added, so the value represents accumulated absorbed deformation work rather than unloading recovery. The final absorbed energies on the shorter-span and longer-span sides are \(E_n\) and \(E_f\), respectively. The absorbed energies per unit span length are
\begin{equation}
  \bar{E}_n=\frac{E_n}{a_n},
  \qquad
  \bar{E}_f=\frac{E_f}{a_f}.
  \label{eq:energy_density_near_far}
\end{equation}
The shorter/longer energy-density ratio and the shorter-span-side energy-density amplification are
\begin{equation}
  R_{\bar{E}}=
  \frac{\bar{E}_n}{\bar{E}_f},
  \qquad
  A_{\bar{E},n}(e,v,f_c)=
  \frac{\bar{E}_n(e,v,f_c)}
  {\bar{E}_n(0,v,f_c)} .
  \label{eq:energy_density_main_indicators}
\end{equation}
The ratio \(R_{\bar{E}}\) compares the absorbed energy per unit length on the shorter-span side with that on the longer-span side in the same case. The amplification \(A_{\bar{E},n}\) shows how many times the shorter-span-side absorbed energy per unit length increases after the impact point moves away from midspan, using the central-impact case with the same impact velocity and concrete strength as the baseline.

Damage and fragmentation are evaluated from selected damage-field snapshots. The snapshots are used to read local crushing below the impact point, damage spreading toward the shorter-span side, tensile-side cracking near the beam bottom, support-adjacent damage, and detached fragment patterns.

Table~\ref{tab:indicators_data_sources} summarizes the response quantities used later for the full-span eccentric-impact cases.

\begin{table}[H]
  \caption{Full-span response quantities used in the eccentric-impact analysis.\label{tab:indicators_data_sources}}
  \begin{adjustwidth}{-\extralength}{0cm}
    \centering
    \tablesize{\footnotesize}
    \begingroup
    \setlength{\tabcolsep}{4pt}
    \renewcommand{\arraystretch}{1.12}
    \begin{tabularx}{\fulllength}{@{}>{\raggedright\arraybackslash}p{0.25\fulllength}>{\raggedright\arraybackslash}p{0.24\fulllength}>{\raggedright\arraybackslash}X@{}}
      \toprule
      \textbf{Response component} & \textbf{Quantity} & \textbf{Response information} \\
      \midrule
      Contact pulse
      & \(F_1\), \(T_p\)
      & Initial contact-force scale and duration of the main contact event \\

      Sectional shear
      & \(A_{V,n}\), \(R_V\)
      & Shorter-span shear increase relative to central impact and comparison with the longer-span side \\

      Sectional moment
      & \(R_M\)
      & Shorter/longer moment comparison for checking whether bending follows the shear trend \\

      Absorbed energy per unit length
      & \(A_{\bar{E},n}\), \(R_{\bar{E}}\)
      & Shorter-span energy increase relative to central impact and comparison with the longer-span side \\

      Damage morphology
      & Damage field and detached fragments
      & Damage localization, spreading, and fragment development \\
      \bottomrule
    \end{tabularx}
    \endgroup
    {\raggedright\footnotesize\noindent\textit{Note:} Quantities with \(A\) compare each eccentric case with the corresponding central-impact case in the same velocity and concrete-strength branch. Quantities with \(R\) compare the shorter-span side with the longer-span side within the same case.\par}
    \tablesize{}
  \end{adjustwidth}
\end{table}

The comparison with short-span reference beams is defined separately in Section~\ref{sec:matched_short_span_reference}.

\subsection{Matched short-span reference cases}
\label{sec:matched_short_span_reference}

\begin{table}[H]
  \caption{Matched short-span reference cases and paired full-span eccentric cases.\label{tab:short_span_reference_matrix}}
  \begin{adjustwidth}{-\extralength}{0cm}
    \centering
    \tablesize{\footnotesize}
    \begin{tabularx}{\fulllength}{LLLLCCCCC}
      \toprule
      \textbf{Ref. level} & \textbf{Ref. case} & \textbf{Paired full-span case} & \textbf{Paired level} & \boldmath{\(v\)} & \boldmath{\(f_c\)} & \textbf{Matched} \boldmath{\(a_n\)} & \boldmath{\(L_S\)} & \textbf{Beam length} \\
      & & & & \textbf{(m/s)} & \textbf{(MPa)} & \textbf{(mm)} & \textbf{(mm)} & \textbf{(mm)} \\
      \midrule
      \(\mathrm{S1}\) & \(\mathrm{S4C35S1}\) & \(\mathrm{S4C35E1}\) & \(\mathrm{E1}\) & \(4\) & \(35\) & \(400\) & \(800\) & \(1000\) \\
      \(\mathrm{S1}\) & \(\mathrm{S4C45S1}\) & \(\mathrm{S4C45E1}\) & \(\mathrm{E1}\) & \(4\) & \(45\) & \(400\) & \(800\) & \(1000\) \\
      \(\mathrm{S2}\) & \(\mathrm{S4C35S2}\) & \(\mathrm{S4C35E2}\) & \(\mathrm{E2}\) & \(4\) & \(35\) & \(300\) & \(600\) & \(800\) \\
      \(\mathrm{S2}\) & \(\mathrm{S4C45S2}\) & \(\mathrm{S4C45E2}\) & \(\mathrm{E2}\) & \(4\) & \(45\) & \(300\) & \(600\) & \(800\) \\
      \(\mathrm{S3}\) & \(\mathrm{S4C35S3}\) & \(\mathrm{S4C35E3}\) & \(\mathrm{E3}\) & \(4\) & \(35\) & \(200\) & \(400\) & \(600\) \\
      \(\mathrm{S3}\) & \(\mathrm{S4C45S3}\) & \(\mathrm{S4C45E3}\) & \(\mathrm{E3}\) & \(4\) & \(45\) & \(200\) & \(400\) & \(600\) \\
      \bottomrule
    \end{tabularx}
    {\raggedright\footnotesize\noindent\textit{Note:} The reference support span is \(L_S=2a_n\), so one half of the symmetric reference beam matches the shorter span of the paired full-span eccentric case.\par}
    \tablesize{}
  \end{adjustwidth}
\end{table}

The matched short-span reference cases test whether the shorter-span side of an eccentric full-span beam can be evaluated as a separate symmetric short-span beam. If the full-span result is close to the reference result, the short-span response can be treated mainly as a local span effect. If the full-span beam gives larger shear, larger bending moment, or more severe damage, the longer-span side and the original full-span boundary condition are still affecting the response on the shorter-span side.

For each eccentricity level \(\mathrm{E}k\), a reference level \(\mathrm{S}k\) is introduced. The reference beam uses the same impact velocity and concrete strength as the paired full-span eccentric case. Its support span \(L_S\) is chosen so that one half of the reference span is equal to the shorter impact-to-support distance \(a_n\) of the paired full-span case:
\[
  L_S = 2a_n .
\]
The same \(100~\mathrm{mm}\) end extension beyond each support is retained, so the total beam length of the reference model is \(L_S+200~\mathrm{mm}\). The reference set is prepared for the \(v=4~\mathrm{m/s}\), \(f_c=35\) and \(45~\mathrm{MPa}\) cases used in the matched comparison. Table~\ref{tab:short_span_reference_matrix} lists the paired cases and geometries.

\begin{figure}[H]
  \centering
  \includegraphics[width=0.92\linewidth]{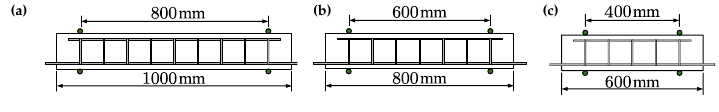}
  \caption{Matched short-span reference geometries: (a) \(\mathrm{S1}\), (b) \(\mathrm{S2}\), and (c) \(\mathrm{S3}\).\label{fig:matched_short_span_reference}}
\end{figure}

The comparison is made with actual shear and bending-moment values in physical units. For pair \(k\), the full-span values are taken from the shorter-span side of the full-span eccentric beam:
\begin{equation}
  V_{\mathrm{full}}^{(k)}
  =
  V_{n,\max}^{>5\mathrm{ms}}(\mathrm{E}k),
  \qquad
  M_{\mathrm{full}}^{(k)}
  =
  M_{n,\max}^{>5\mathrm{ms}}(\mathrm{E}k).
  \label{eq:matched_full_span_demand}
\end{equation}
The reference values are taken from the matching half of the symmetric short-span reference beam:
\begin{equation}
  V_{\mathrm{ref}}^{(k)}
  =
  V_{S,\max}^{>5\mathrm{ms}}(\mathrm{S}k),
  \qquad
  M_{\mathrm{ref}}^{(k)}
  =
  M_{S,\max}^{>5\mathrm{ms}}(\mathrm{S}k).
  \label{eq:matched_reference_demand}
\end{equation}
Here, \(V_{\mathrm{full}}^{(k)}\) and \(M_{\mathrm{full}}^{(k)}\) are the largest post-\(5~\mathrm{ms}\) shear and bending-moment envelope values on the shorter-span side of the full-span eccentric case. \(V_{\mathrm{ref}}^{(k)}\) and \(M_{\mathrm{ref}}^{(k)}\) are the corresponding largest post-\(5~\mathrm{ms}\) values in the matched half of the reference case.

The difference between the full-span case and the matched reference is reported as
\begin{equation}
  \Delta V^{(k)}
  =
  V_{\mathrm{full}}^{(k)}-V_{\mathrm{ref}}^{(k)},
  \qquad
  \Delta M^{(k)}
  =
  M_{\mathrm{full}}^{(k)}-M_{\mathrm{ref}}^{(k)}.
  \label{eq:matched_absolute_differences}
\end{equation}
A positive difference means that the full-span eccentric beam produces a larger value than the symmetric short-span reference. These differences are simple subtractions of the two actual values.

When envelope profiles are compared, the local coordinate \(s\) starts from the impact section and increases toward the compared support. If the reference half span is drawn on the opposite side, its profile is mirrored onto the same positive \(s\) direction. This lets the two local-span profiles be read on the same coordinate.

The matched-reference comparison uses three types of output. The first is the peak shear and bending-moment demand: \(V_{\mathrm{full}}\) and \(M_{\mathrm{full}}\) are taken from the shorter-span side of the full-span eccentric beam, and \(V_{\mathrm{ref}}\) and \(M_{\mathrm{ref}}\) are taken from the matched half span of the symmetric short-span reference beam. Their differences, \(\Delta V\) and \(\Delta M\), show whether the full-span eccentric beam develops larger local demand than the reference beam with the same shorter span.

The second output is the local envelope profile plotted along \(s\), which shows where the shear and bending-moment differences occur along the short span. The third output is the paired final damage field, which shows whether the difference in sectional demand is accompanied by more connected cracking, crushing, or detached fragments in the full-span eccentric beam.

\section{Results and Discussion}

\subsection{Full-span contact-force overview}

\begin{figure}[H]
  \centering
  \includegraphics[width=13.5cm]{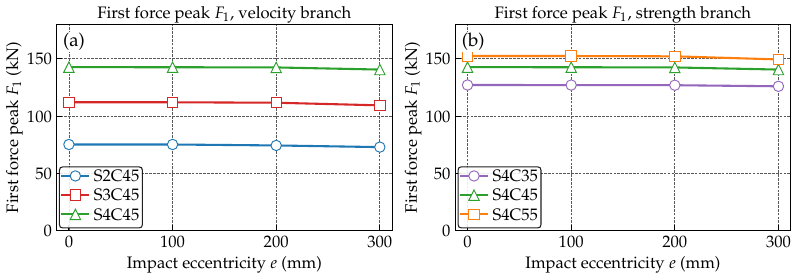}
  \caption{Full-span first force peak \(F_1\) versus impact eccentricity. Panel (a) shows the velocity branch at \(f_c=45~\mathrm{MPa}\), and panel (b) shows the concrete-strength branch at \(v=4~\mathrm{m/s}\).\label{fig:global_first_force_peak}}
\end{figure}

\begin{figure}[H]
  \centering
  \includegraphics[width=13.5cm]{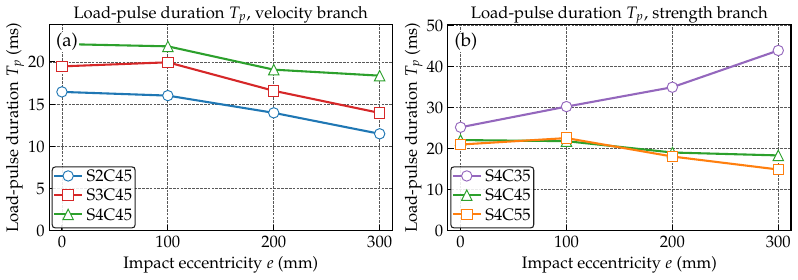}
  \caption{Full-span load-pulse duration \(T_p\) versus impact eccentricity. Panel (a) shows the velocity branch at \(f_c=45~\mathrm{MPa}\), and panel (b) shows the concrete-strength branch at \(v=4~\mathrm{m/s}\).\label{fig:global_pulse_duration}}
\end{figure}

Table~\ref{tab:global_response_summary} lists two contact-force quantities for the full-span cases: the first force peak and the duration of the main contact pulse. Figures~\ref{fig:global_first_force_peak} and~\ref{fig:global_pulse_duration} show how these two values change as the impact point moves away from midspan.

\begin{table}[H]
  \caption{Full-span contact-force response quantities.\label{tab:global_response_summary}}
  \centering
  \tablesize{\footnotesize}
  \setlength{\tabcolsep}{2pt}
  \begin{tabularx}{\linewidth}{@{}>{\raggedright\arraybackslash}p{0.15\linewidth}>{\centering\arraybackslash}p{0.06\linewidth}>{\centering\arraybackslash}p{0.07\linewidth}>{\centering\arraybackslash}p{0.06\linewidth}CC@{}}
    \toprule
    \textbf{Case} & \boldmath{\(v\)} & \boldmath{\(f_c\)} & \boldmath{\(e\)} & \mbox{\textbf{First force peak} \boldmath{\(F_1\)}} & \mbox{\textbf{Load-pulse duration} \boldmath{\(T_p\)}} \\
    & \textbf{(m/s)} & \textbf{(MPa)} & \textbf{(mm)} & \textbf{(kN)} & \textbf{(ms)} \\
    \midrule
    S2C45E0 & 2 & 45 & 0 & 75.3 & 16.45 \\
    S2C45E1 & 2 & 45 & 100 & 75.3 & 16.01 \\
    S2C45E2 & 2 & 45 & 200 & 74.5 & 13.97 \\
    S2C45E3 & 2 & 45 & 300 & 73.0 & 11.49 \\
    S3C45E0 & 3 & 45 & 0 & 112.1 & 19.47 \\
    S3C45E1 & 3 & 45 & 100 & 112.1 & 19.95 \\
    S3C45E2 & 3 & 45 & 200 & 111.7 & 16.58 \\
    S3C45E3 & 3 & 45 & 300 & 109.4 & 13.96 \\
    S4C45E0 & 4 & 45 & 0 & 142.7 & 22.11 \\
    S4C45E1 & 4 & 45 & 100 & 142.7 & 21.84 \\
    S4C45E2 & 4 & 45 & 200 & 142.5 & 19.08 \\
    S4C45E3 & 4 & 45 & 300 & 140.6 & 18.37 \\
    S4C35E0 & 4 & 35 & 0 & 127.2 & 25.22 \\
    S4C35E1 & 4 & 35 & 100 & 127.1 & 30.25 \\
    S4C35E2 & 4 & 35 & 200 & 127.0 & 34.97 \\
    S4C35E3 & 4 & 35 & 300 & 126.0 & 43.91 \\
    S4C55E0 & 4 & 55 & 0 & 152.5 & 21.02 \\
    S4C55E1 & 4 & 55 & 100 & 152.4 & 22.59 \\
    S4C55E2 & 4 & 55 & 200 & 152.2 & 18.10 \\
    S4C55E3 & 4 & 55 & 300 & 149.4 & 14.94 \\
    \bottomrule
  \end{tabularx}
  {\raggedright\footnotesize\noindent\textit{Note:} The first force peak \(F_1\) gives the initial contact-force scale. The load-pulse duration \(T_p\) is the duration for which the impactor--beam contact force remains present during the main impact event.\par}
  \tablesize{}
\end{table}

The first force peak changes little as the impact point moves away from midspan. From central impact to the \(300~\mathrm{mm}\) eccentric case, the first force peak decreases by \(2.3~\mathrm{kN}\), \(2.7~\mathrm{kN}\), and \(2.1~\mathrm{kN}\) for the \(\mathrm{S2C45}\), \(\mathrm{S3C45}\), and \(\mathrm{S4C45}\) velocity groups. In the strength branch, the corresponding decreases are \(1.2~\mathrm{kN}\), \(2.1~\mathrm{kN}\), and \(3.1~\mathrm{kN}\) for \(\mathrm{S4C35}\), \(\mathrm{S4C45}\), and \(\mathrm{S4C55}\). These small changes indicate that the later redistribution of shear, absorbed energy, and damage arises mainly from the subsequent load-transfer process rather than from a larger first force peak.

In the velocity branch, the main contact pulse becomes shorter as the impact point moves farther from midspan. The \(\mathrm{S2C45}\) cases decrease from \(16.45~\mathrm{ms}\) at central impact to \(11.49~\mathrm{ms}\) at the \(300~\mathrm{mm}\) eccentric case. The \(\mathrm{S3C45}\) cases decrease from \(19.47~\mathrm{ms}\) to \(13.96~\mathrm{ms}\), after a small increase at the \(100~\mathrm{mm}\) eccentric case. The \(\mathrm{S4C45}\) cases decrease from \(22.11~\mathrm{ms}\) to \(18.37~\mathrm{ms}\). Thus, in the velocity branch, larger eccentricity shortens the main contact pulse.

The concrete-strength branch gives a different pattern. The low-strength \(\mathrm{S4C35}\) cases keep contact for longer as eccentricity increases, with the pulse duration increasing from \(25.22~\mathrm{ms}\) at central impact to \(43.91~\mathrm{ms}\) at the \(300~\mathrm{mm}\) eccentric case. By contrast, the \(\mathrm{S4C45}\) cases decrease from \(22.11~\mathrm{ms}\) to \(18.37~\mathrm{ms}\), and the \(\mathrm{S4C55}\) cases decrease from \(21.02~\mathrm{ms}\) to \(14.94~\mathrm{ms}\) after a small increase at \(100~\mathrm{mm}\).

The strength branch separates contact duration from force magnitude. The low-strength \(\mathrm{S4C35}\) branch does not develop a larger first force peak, but it maintains contact for a much longer time as eccentricity increases. The trend is consistent with longer local crushing, separation, and fragment interaction after the initial contact. The contact-force results give the time-scale context for the later sectional-demand, absorbed-energy, and damage analyses.

\subsection{Sectional shear and bending-moment redistribution}

\begin{figure}[H]
  \centering
  \makebox[\linewidth][c]{\includegraphics[width=13.5cm]{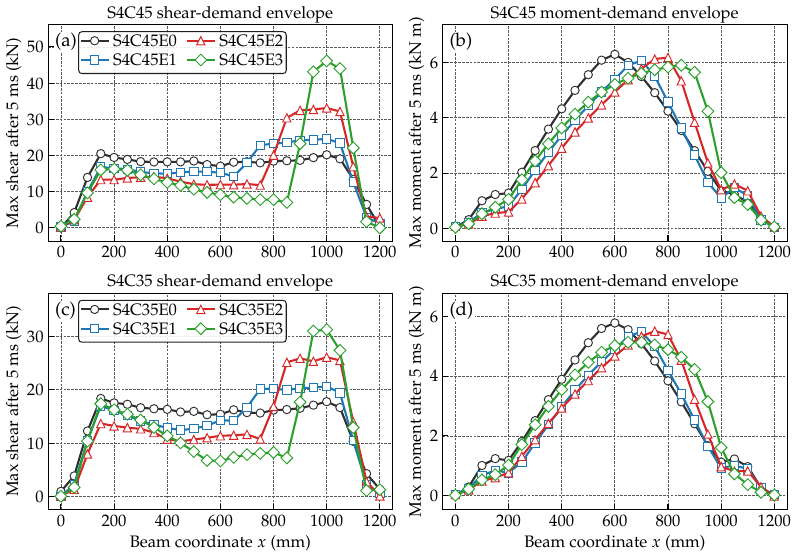}}
  \caption{Sectional shear and bending-moment demand envelopes for representative full-span eccentric cases. Panels (a,b) show the \(\mathrm{S4C45}\) group, and panels (c,d) show the \(\mathrm{S4C35}\) group. Values are post-\(5~\mathrm{ms}\) maximum absolute section resultants along the beam coordinate.\label{fig:sectional_envelopes}}
\end{figure}

Figure~\ref{fig:sectional_envelopes} shows where the largest shear and bending-moment values appear along the beam after the early contact shock. These curves are envelope curves rather than snapshots at one time. At each section position, the plotted value is the largest absolute shear or bending moment recorded after \(5~\mathrm{ms}\). This envelope view shows where the beam may experience its maximum sectional demand during the impact response.

The sectional plots focus on the \(4~\mathrm{m/s}\) cases because these cases show the clearest side-to-side redistribution and also allow the concrete-strength comparison.

\begin{table}[H]
  \caption{Sectional demand envelopes and shorter/longer comparisons for selected \(v=4~\mathrm{m/s}\) full-span cases.\label{tab:sectional_demand_summary}}
  \begin{adjustwidth}{-\extralength}{0cm}
    \centering
    \tablesize{\scriptsize}
    \setlength{\tabcolsep}{2pt}
    \begin{tabularx}{\fulllength}{LCCCCCCC}
      \toprule
      \textbf{Case} & \shortstack[c]{\textbf{Shear increase}\\\textbf{vs. E0}\\\boldmath{\(A_{V,n}\)}} & \shortstack[c]{\textbf{Shorter-span}\\\textbf{shear}\\\boldmath{\(V_n\)}} & \shortstack[c]{\textbf{Longer-span}\\\textbf{shear}\\\boldmath{\(V_f\)}} & \shortstack[c]{\textbf{Shorter/longer}\\\textbf{shear}\\\boldmath{\(R_V\)}} & \shortstack[c]{\textbf{Shorter-span}\\\textbf{moment}\\\boldmath{\(M_n\)}} & \shortstack[c]{\textbf{Longer-span}\\\textbf{moment}\\\boldmath{\(M_f\)}} & \shortstack[c]{\textbf{Shorter/longer}\\\textbf{moment}\\\boldmath{\(R_M\)}} \\
      & & \textbf{(kN)} & \textbf{(kN)} & & \textbf{(kN m)} & \textbf{(kN m)} & \\
      \midrule
      S4C35E0 & 1.00 & 18.4 & 19.2 & 0.96 & 5.75 & 5.80 & 0.99 \\
      S4C35E1 & 1.15 & 21.1 & 17.6 & 1.20 & 5.40 & 5.56 & 0.97 \\
      S4C35E2 & 1.47 & 27.1 & 14.0 & 1.93 & 5.22 & 5.56 & 0.94 \\
      S4C35E3 & 1.77 & 32.5 & 18.2 & 1.79 & 3.95 & 5.15 & 0.77 \\
      S4C45E0 & 1.00 & 20.9 & 21.5 & 0.97 & 6.23 & 6.34 & 0.98 \\
      S4C45E1 & 1.21 & 25.3 & 17.4 & 1.45 & 5.97 & 6.16 & 0.97 \\
      S4C45E2 & 1.62 & 33.8 & 14.2 & 2.38 & 6.07 & 6.27 & 0.97 \\
      S4C45E3 & 2.23 & 46.5 & 16.5 & 2.82 & 5.35 & 5.92 & 0.90 \\
      S4C55E0 & 1.00 & 24.0 & 24.4 & 0.98 & 6.65 & 6.78 & 0.98 \\
      S4C55E1 & 1.08 & 25.8 & 17.8 & 1.45 & 5.93 & 6.25 & 0.95 \\
      S4C55E2 & 1.50 & 35.9 & 14.9 & 2.41 & 6.35 & 6.56 & 0.97 \\
      S4C55E3 & 2.14 & 51.4 & 16.6 & 3.10 & 5.87 & 6.56 & 0.90 \\
      \bottomrule
    \end{tabularx}
    {\raggedright\footnotesize\noindent\textit{Note:} The shear and moment values are post-\(5~\mathrm{ms}\) maximum absolute envelope values. For the central-impact rows \(\mathrm{E0}\), the two half spans are equal; the shorter-span and longer-span columns use the same side assignment as the corresponding eccentric cases in each strength branch. \(A_{V,n}\) compares the shorter-span shear with the corresponding central-impact value. \(R_V\) and \(R_M\) compare the shorter-span and longer-span values within the same case.\par}
    \tablesize{}
  \end{adjustwidth}
\end{table}

For the \(\mathrm{S4C45}\) group, Table~\ref{tab:sectional_demand_summary} shows a clear shift of shear toward the shorter-span side. At the \(300~\mathrm{mm}\) eccentricity level, the shear on the shorter-span side becomes \(2.23\) times the central-impact value. In the same case, the shorter-span shear is \(2.82\) times the longer-span shear. The bending moment does not concentrate in the same way. At the largest eccentricity, the shorter-span bending moment is \(0.90\) times the longer-span value, so bending remains slightly larger on the longer-span side. This contrast indicates that the eccentric response is governed mainly by shear redistribution. The sectional forces do not increase uniformly.

The concrete-strength branch gives the same main contrast. At the \(300~\mathrm{mm}\) eccentricity level, the shorter-span shear is \(1.77\), \(2.23\), and \(2.14\) times the corresponding central-impact value for \(\mathrm{S4C35}\), \(\mathrm{S4C45}\), and \(\mathrm{S4C55}\), respectively. In the same three cases, the shorter-span shear is \(1.79\), \(2.82\), and \(3.10\) times the longer-span shear. The bending-moment comparison is much weaker, with the shorter-span moment remaining between \(0.77\) and \(0.90\) times the longer-span moment. These values show a shear-dominated redistribution toward the shorter-span side, while bending remains a broader beam-scale response.

The \(\mathrm{S4C35}\) group also shows why the shorter-span shear should be read together with the shorter/longer shear comparison. From the \(200~\mathrm{mm}\) to the \(300~\mathrm{mm}\) eccentric case, the shorter-span shear increases from \(27.1\) to \(32.5~\mathrm{kN}\). At the same time, the longer-span shear also rises from \(14.0\) to \(18.2~\mathrm{kN}\). Because both sides increase, the shorter-to-longer shear comparison decreases slightly, from \(1.93\) to \(1.79\), even though the shorter-span shear itself still increases. This means that severe low-strength damage activates a broader dynamic response, while the shorter-span side remains the main shear-transfer path.

The envelope profiles in Figure~\ref{fig:sectional_envelopes} show the spatial form behind the values in Table~\ref{tab:sectional_demand_summary}. In the \(\mathrm{S4C45}\) group, the shear envelope becomes increasingly concentrated along the shorter-span transfer route as eccentricity increases, with the largest separation appearing at the \(300~\mathrm{mm}\) eccentricity level. The bending-moment envelopes do not show the same concentration. Their peaks remain distributed over a wider beam-scale region. The \(\mathrm{S4C35}\) group shows a broader damaged-response pattern: at the largest eccentricity, both the shorter-span shear and the longer-span shear increase, which is consistent with the fragment and damage observations discussed later.

\subsection{Shorter-span absorbed-energy density localization}

\begin{figure}[H]
  \centering
  \makebox[\linewidth][c]{\includegraphics[width=13.5cm]{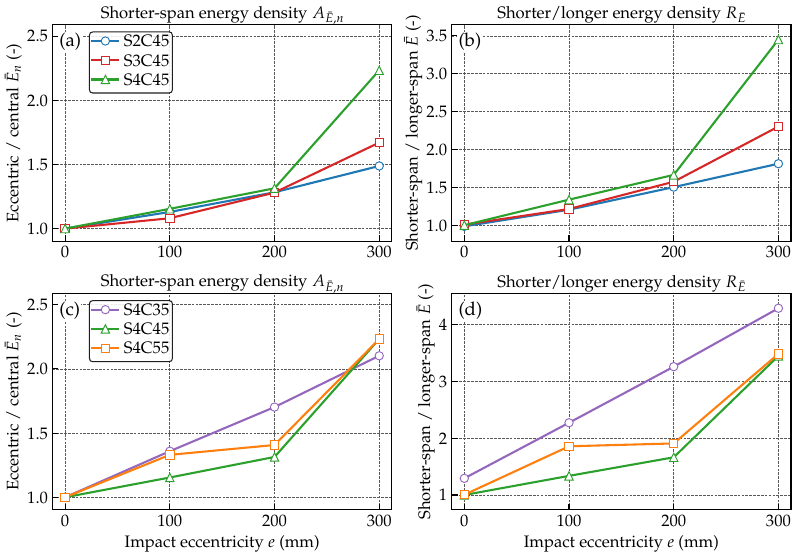}}
  \caption{Shorter-span absorbed-energy density trends. Panels (a,b) show the velocity branch at \(f_c=45~\mathrm{MPa}\), and panels (c,d) show the strength branch at \(v=4~\mathrm{m/s}\). The left column gives the increase relative to the corresponding central-impact case, and the right column gives the shorter-span/longer-span comparison.\label{fig:energy_density_trends}}
\end{figure}

Figure~\ref{fig:energy_density_trends} compares absorbed energy per unit span length, because the two sides of an eccentric beam have different lengths. The left-column plots show how many times the shorter-span-side absorbed energy per unit length increases after the impact point moves away from midspan. The right-column plots compare the absorbed energy per unit length on the shorter-span side with that on the longer-span side in the same case.

In the velocity branch, both energy comparisons increase as the impact point moves farther from midspan. At the \(300~\mathrm{mm}\) eccentricity level, the absorbed energy per unit length on the shorter-span side is \(1.49\), \(1.67\), and \(2.23\) times the corresponding central-impact value for \(\mathrm{S2C45}\), \(\mathrm{S3C45}\), and \(\mathrm{S4C45}\), respectively. In the same cases, the shorter-span side carries \(1.81\), \(2.30\), and \(3.45\) times the absorbed energy per unit length of the longer-span side.

In the concrete-strength branch, the increase relative to the central-impact case remains close to \(2.1\) to \(2.23\) at the \(300~\mathrm{mm}\) eccentricity level. The side-to-side comparison separates the strength groups more clearly. For \(\mathrm{S4C35E3}\), the shorter-span side carries \(4.29\) times the absorbed energy per unit length of the longer-span side. The corresponding values are \(3.45\) for \(\mathrm{S4C45E3}\) and \(3.49\) for \(\mathrm{S4C55E3}\).

The energy result gives the deformation-work side of the same trend observed in the shear envelopes. The shorter-span side carries more shear and also absorbs more deformation work per unit length. The low-strength branch is especially sensitive in the side-to-side energy comparison, which explains why the \(\mathrm{S4C35}\) damage fields later show more visible detached fragments.

\subsection{Matched short-span reference comparison}

The matched short-span cases test whether the shorter-span side of the eccentric full-span beam can be represented by a separate symmetric short-span beam. Each reference beam has one half span equal to the shorter impact-to-support distance of its paired full-span case. The comparison uses shear and bending-moment values in physical units, so the two beams can be read directly.

The available matched-reference simulations cover the \(4~\mathrm{m/s}\), \(35\) and \(45~\mathrm{MPa}\) cases. The \(55~\mathrm{MPa}\) full-span cases are retained in the strength comparison, but paired short-span references were not prepared for that strength level.

Table~\ref{tab:matched_short_span_summary} compares the peak shear and bending-moment demand of each full-span eccentric case with its matched short-span reference. The table reports the actual values and their arithmetic differences. The shear and bending-moment trends are considered separately because the two resultants respond differently to the matched-span condition.

\begin{table}[H]
  \caption{Peak shear and bending-moment demand in full-span eccentric cases and matched short-span references.\label{tab:matched_short_span_summary}}
  \begin{adjustwidth}{-\extralength}{0cm}
    \centering
    \tablesize{\footnotesize}
    \setlength{\tabcolsep}{3pt}
    \begin{tabularx}{\fulllength}{LLCCCCCCC}
      \toprule
      \textbf{Full-span case} & \textbf{Reference case} & \boldmath{\(a_n\)} & \boldmath{\(V_{\mathrm{full}}\)} & \boldmath{\(V_{\mathrm{ref}}\)} & \boldmath{\(\Delta V\)} & \boldmath{\(M_{\mathrm{full}}\)} & \boldmath{\(M_{\mathrm{ref}}\)} & \boldmath{\(\Delta M\)} \\
      & & \textbf{(mm)} & \textbf{(kN)} & \textbf{(kN)} & \textbf{(kN)} & \textbf{(kN m)} & \textbf{(kN m)} & \textbf{(kN m)} \\
      \midrule
      \(\mathrm{S4C35E1}\) & \(\mathrm{S4C35S1}\) & \(400\) & \(21.1\) & \(21.4\) & \(-0.3\) & \(5.40\) & \(5.28\) & \(0.12\) \\
      \(\mathrm{S4C35E2}\) & \(\mathrm{S4C35S2}\) & \(300\) & \(27.1\) & \(24.1\) & \(3.0\) & \(5.22\) & \(5.19\) & \(0.03\) \\
      \(\mathrm{S4C35E3}\) & \(\mathrm{S4C35S3}\) & \(200\) & \(32.5\) & \(21.4\) & \(11.1\) & \(3.95\) & \(2.93\) & \(1.02\) \\
      \(\mathrm{S4C45E1}\) & \(\mathrm{S4C45S1}\) & \(400\) & \(25.3\) & \(24.2\) & \(1.1\) & \(5.97\) & \(5.87\) & \(0.10\) \\
      \(\mathrm{S4C45E2}\) & \(\mathrm{S4C45S2}\) & \(300\) & \(33.8\) & \(29.8\) & \(4.0\) & \(6.07\) & \(6.00\) & \(0.07\) \\
      \(\mathrm{S4C45E3}\) & \(\mathrm{S4C45S3}\) & \(200\) & \(46.5\) & \(28.1\) & \(18.4\) & \(5.35\) & \(3.88\) & \(1.47\) \\
      \bottomrule
    \end{tabularx}
    {\raggedright\footnotesize\noindent\textit{Note:} \(a_n\) is the shorter impact-to-support distance shared by each full-span eccentric case and its matched short-span reference. \(V_{\mathrm{full}}\) and \(M_{\mathrm{full}}\) are the shorter-span-side peak demand envelopes of the full-span eccentric case after \(t_s=5~\mathrm{ms}\). \(V_{\mathrm{ref}}\) and \(M_{\mathrm{ref}}\) are the corresponding peak envelopes in the matched half span of the symmetric short-span reference. The differences are calculated as \(\Delta V=V_{\mathrm{full}}-V_{\mathrm{ref}}\) and \(\Delta M=M_{\mathrm{full}}-M_{\mathrm{ref}}\) from the displayed values. Positive values indicate larger demand in the full-span eccentric case.\par}
    \tablesize{}
  \end{adjustwidth}
\end{table}

\begin{figure}[H]
  \centering
  \makebox[\linewidth][c]{\includegraphics[width=13.5cm]{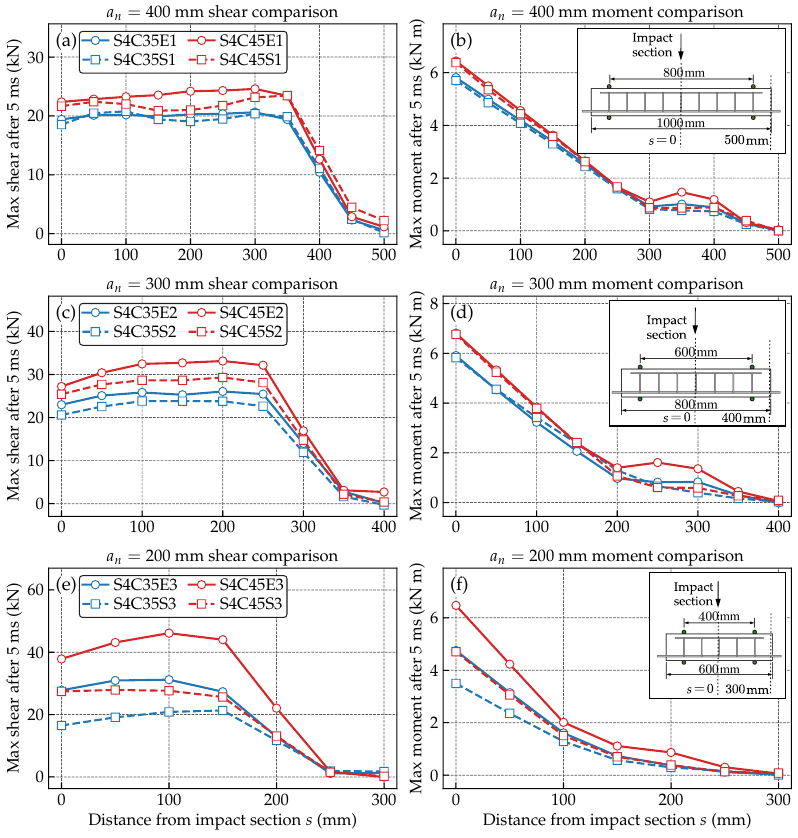}}
  \caption{Paired local-span shear and bending-moment envelopes for full-span eccentric cases and matched short-span references. Rows correspond to \(a_n=400\), \(300\), and \(200~\mathrm{mm}\). The short segment beyond the paired support is retained to show the full local profile.\label{fig:matched_short_span_profiles}}
\end{figure}

For the \(\mathrm{S4C45}\) group, the \(200~\mathrm{mm}\) case and especially the \(300~\mathrm{mm}\) case show the clearest increase in shorter-span shear. In \(\mathrm{S4C45E3}\), the full-span eccentric beam reaches \(46.5~\mathrm{kN}\) in shorter-span shear, while the matched short-span reference reaches \(28.1~\mathrm{kN}\). The difference is \(18.4~\mathrm{kN}\). The bending-moment difference is smaller in relative importance: the full-span value is \(5.35~\mathrm{kN\,m}\), the reference value is \(3.88~\mathrm{kN\,m}\), and the difference is \(1.47~\mathrm{kN\,m}\). The \(100~\mathrm{mm}\) case shows only a marginal shear increase, and the bending-moment differences remain secondary compared with the shear change.

For the \(\mathrm{S4C35}\) group, the shorter-span shear is close to the reference at \(100~\mathrm{mm}\), increases at \(200~\mathrm{mm}\), and becomes clearly larger at \(300~\mathrm{mm}\). The \(100~\mathrm{mm}\) pair gives \(21.1~\mathrm{kN}\) for the full-span eccentric beam and \(21.4~\mathrm{kN}\) for the reference. The \(200~\mathrm{mm}\) pair gives \(27.1~\mathrm{kN}\) and \(24.1~\mathrm{kN}\). The \(300~\mathrm{mm}\) pair gives \(32.5~\mathrm{kN}\) and \(21.4~\mathrm{kN}\). The bending-moment difference is local and secondary. The \(100\) and \(200~\mathrm{mm}\) cases show only marginal moment differences, while the \(300~\mathrm{mm}\) case shows a local moment increase that is still weaker than the shear separation.

Table~\ref{tab:matched_short_span_summary} gives the largest values, and Figure~\ref{fig:matched_short_span_profiles} shows where the differences occur along the short side. For the \(400~\mathrm{mm}\) short span, the full-span and reference curves remain close, which agrees with the small peak differences in the table. For the \(300~\mathrm{mm}\) short span, the full-span eccentric cases begin to stay above the references over much of the local span. For the \(200~\mathrm{mm}\) short span, the separation is strongest, especially for \(\mathrm{S4C45E3}\), where the full-span shear curve stays well above the reference before dropping near the support-end region. The moment curves show a more local difference near the impact side and do not separate over the span as clearly as the shear curves.

The matched-reference result is more than a short-span strength check. If the eccentric full-span response were controlled only by the isolated short span, the full-span and reference curves would stay close after matching the short span length. The increasing separation from \(100\) to \(300~\mathrm{mm}\) eccentricity, especially in shear, shows that the longer side and the full-span deformation shape still affect the demand carried through the shorter-span side.

\subsection{Damage asymmetry and fragmentation pattern}

\begin{figure}[H]
  \begin{adjustwidth}{-\extralength}{0cm}
    \centering
    \includegraphics[width=\fulllength]{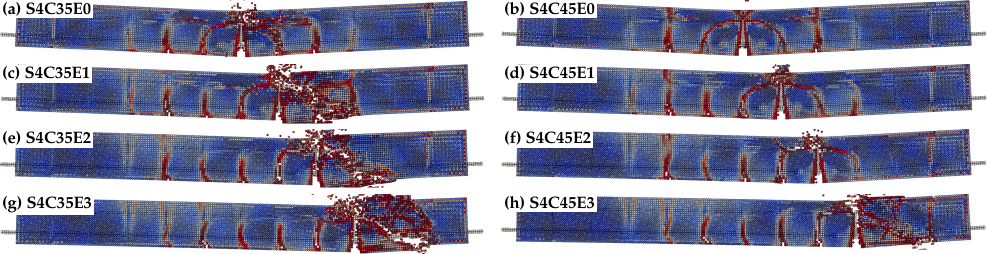}
    \caption{Final-frame damage fields for the \(\mathrm{S4C35E}\) and \(\mathrm{S4C45E}\) groups. Rows correspond to \(\mathrm{E0}\)--\(\mathrm{E3}\), and columns compare concrete strengths.\label{fig:damage_asymmetry_s4c35_s4c45}}
  \end{adjustwidth}
\end{figure}

Figure~\ref{fig:damage_asymmetry_s4c35_s4c45} shows the final damage fields for the \(\mathrm{S4C35E}\) and \(\mathrm{S4C45E}\) groups. The panels are arranged by eccentricity level, with the \(\mathrm{S4C35E}\) cases in the left column and the \(\mathrm{S4C45E}\) cases in the right column. The comparison focuses on where the damage localizes, how far it spreads from the impact region, and whether detached fragments develop along the shorter-span transfer path.

At central impact, the main damage remains centered around the impact section. As the impact point moves away from midspan, the dominant damage region shifts toward the shorter-span side. This shift is clearer in the low-strength \(\mathrm{S4C35}\) group. At the largest eccentricity, \(\mathrm{S4C35E3}\) develops a broad detached and splashed particle region near the shortened transfer path, while \(\mathrm{S4C45E3}\) shows a more bounded damage zone. This visual difference is consistent with the longer contact duration of the low-strength branch and with the stronger absorbed-energy concentration reported above.

\begin{figure}[H]
  \centering
  \includegraphics[width=\linewidth]{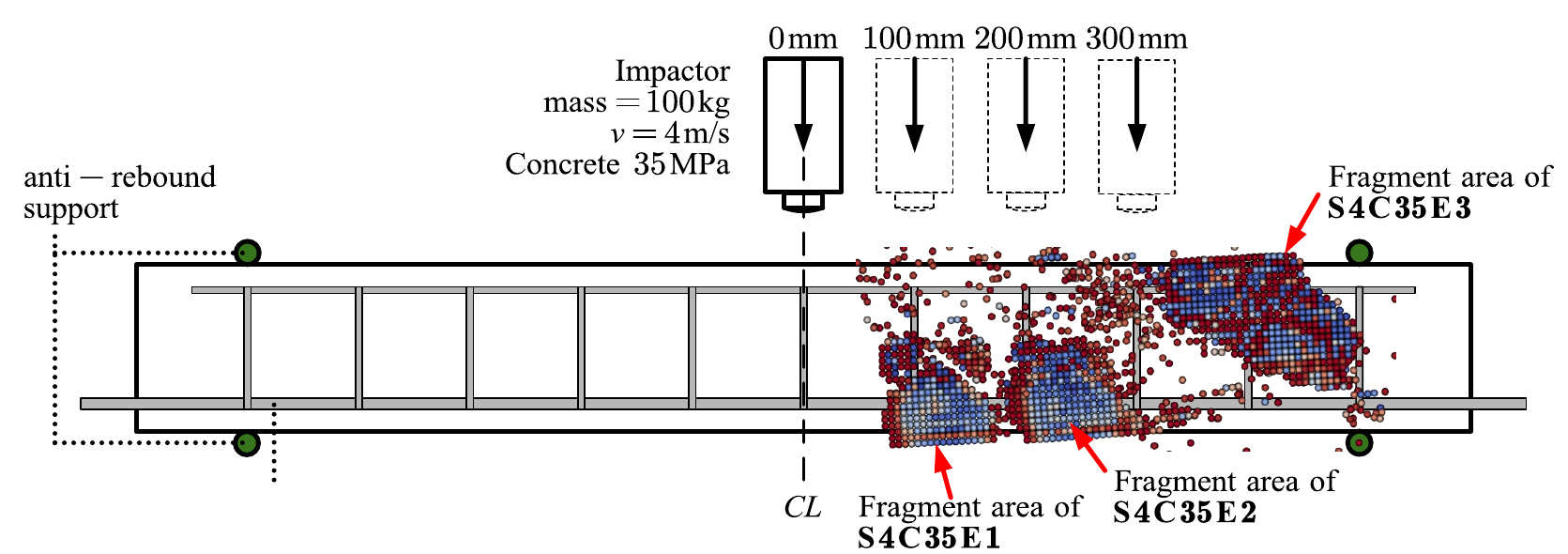}
  \caption{Velocity-filtered detached fragment areas in the \(\mathrm{S4C35}\) eccentric cases. Highlighted regions mark concrete blocks moving independently from the main beam rebound.\label{fig:fragment_area_compare}}
\end{figure}

Figure~\ref{fig:fragment_area_compare} gives a closer view of detached fragments in the \(\mathrm{S4C35}\) eccentric cases. The highlighted regions mark concrete pieces that separate from the main beam body after local crushing and rebound. As eccentricity increases, the detached region becomes more evident and shifts toward the shorter-span transfer path. The trend is consistent with the rise in shorter-span shear and absorbed energy per unit length.

The final morphology check compares the full-span eccentric cases with their matched short-span references.

\begin{figure}[H]
  \begin{adjustwidth}{-\extralength}{0cm}
    \makebox[\fulllength][r]{\includegraphics[width=\fulllength]{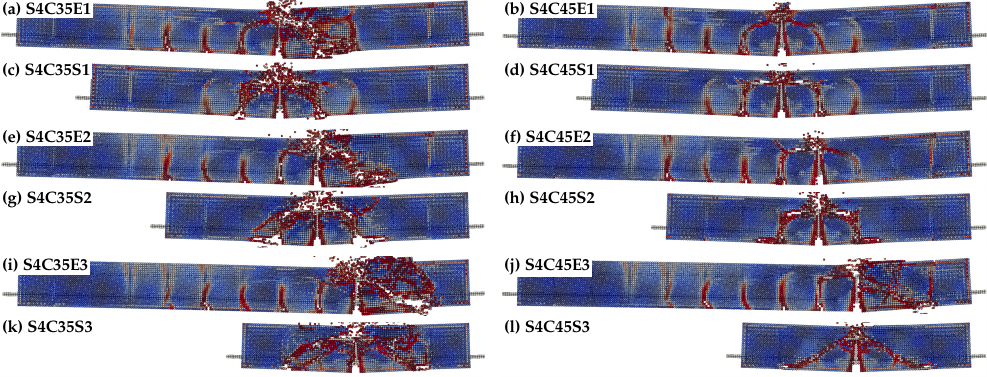}}
    \caption{Paired damage-field comparison between full-span eccentric cases and matched short-span references. Each pair compares the full-span eccentric case with the reference beam having the same shorter span. The figure shows the final-frame damage morphology for each paired case.\label{fig:matched_reference_damage_comparison}}
  \end{adjustwidth}
\end{figure}

Figure~\ref{fig:matched_reference_damage_comparison} shows the damage counterpart of the matched-reference comparison. At the \(100~\mathrm{mm}\) eccentricity level, the full-span eccentric and matched-reference damage fields remain broadly comparable, with visible differences mainly around the impact region. This agrees with the comparable or marginal peak-demand differences in Table~\ref{tab:matched_short_span_summary}. At \(200~\mathrm{mm}\), the full-span eccentric cases show more extension toward the shorter-span transfer route, while the matched references remain more compact around the local impact span. At \(300~\mathrm{mm}\), the contrast becomes clearer. The full-span eccentric cases develop a more connected shorter-span-side damage zone and more visible detached material than the corresponding short-span references. The larger and more connected damage zones in the full-span cases show that the longer side and the full-span deformation shape still affect damage development, even when the shorter span length is matched.

\subsection{Mechanism of eccentricity-induced load-path redistribution}

The response is controlled by a load-path change associated with the shorter impact-to-support distance. The first force peak changes little as eccentricity increases, so the later redistribution is not driven mainly by a larger initial contact force. The contact duration gives the time-scale context. In the velocity branch, the pulse generally becomes shorter as eccentricity increases. In the low-strength \(\mathrm{S4C35}\) branch, however, the contact duration increases markedly, which is consistent with longer local crushing, separation, and fragment interaction.

The sectional results identify the spatial redistribution. Moving the impact point away from midspan shortens one side, and this shorter side provides a more direct force-transfer route to the closer support. The increase in shorter-span shear relative to both the central-impact case and the longer-span side reflects this change. The bending response does not follow the same trend. The shorter-span bending moment remains close to or lower than the longer-span value, and the moment envelopes keep a broader beam-scale shape. The load-path change is therefore shear-dominated rather than a uniform increase of all sectional demands.

The energy result shows the same mechanism from the work-absorption side. Because the two sides have different lengths, absorbed energy per unit length is the relevant comparison. After eccentricity is introduced, the shorter-span side absorbs more deformation work per unit length than in the corresponding central-impact case. It also absorbs more work per unit length than the longer-span side in the same eccentric case. This effect is strongest in the low-strength branch, where the damage fields show broader detached material.

The matched-reference and damage comparisons show how the local shorter span interacts with the remaining beam length. In the more eccentric pairs, the full-span eccentric beams develop larger shorter-span shear than the symmetric short-span references with the same local span length. The envelope profiles show that the difference is spatial, not just a single peak change. At \(200~\mathrm{mm}\) and especially \(300~\mathrm{mm}\) eccentricity, the full-span shear curve separates from the reference over much of the local span. The damage fields show the same tendency: the full-span eccentric beams develop more connected shorter-span-side damage and more visible detached fragments than their matched references. The eccentric full-span response is therefore influenced by more than the local shorter span. The longer side and the global deformation shape still participate in force transfer and damage development.


\section{Conclusions}

This study examined how impact eccentricity changes force transfer and damage development in RC beams using a validated coupled SPH--FEM model. The main conclusions are as follows.

\begin{enumerate}
  \item The coupled SPH--FEM model reproduced the central-impact benchmark response with acceptable accuracy. The validation errors were \(0.1\)--\(2.6\%\) for maximum displacement, \(1.5\)--\(5.7\%\) for the first force peak, and \(3.0\)--\(6.7\%\) for load-pulse duration. The model also captured the main force-history, absorbed-energy, and damage-pattern features. This agreement provides the basis for using the model to examine eccentric-impact response.

  \item Impact eccentricity changed the load-transfer process more clearly than the initial contact-force scale. From central impact to the largest eccentricity, the first force peak changed only slightly, with the largest decrease remaining about \(3.1~\mathrm{kN}\) among the analysed branches. In contrast, the contact duration showed a clear strength-dependent response. In the low-strength branch, the main contact pulse increased from \(25.22~\mathrm{ms}\) to \(43.91~\mathrm{ms}\), reflecting longer local crushing, separation, and fragment interaction.

  \item Eccentric impact redirected shear demand toward the shorter-span side. At the largest eccentricity, the shorter-span shear reached up to \(2.23\) times the corresponding central-impact value and up to \(3.10\) times the longer-span-side value. The bending response did not follow the same trend; the shorter/longer moment ratio remained between \(0.77\) and \(0.90\) at the largest eccentricity. The load-path change is therefore governed mainly by shear redistribution rather than by a uniform increase of sectional demand.

  \item The absorbed-energy results confirmed the same mechanism from the deformation-work side. After normalizing by span length, absorbed energy concentrated on the shorter-span side. At the largest eccentricity, the shorter-span side absorbed up to \(4.29\) times the energy per unit length of the longer-span side. This energy localization explains why the low-strength cases developed stronger detached-fragment morphology.

  \item The matched short-span comparison showed that the eccentric full-span response cannot be represented by an isolated symmetric short-span beam alone. The full-span eccentric cases developed up to \(18.4~\mathrm{kN}\) higher shorter-span shear and \(1.47~\mathrm{kN\,m}\) higher bending moment than the matched references. The shear difference was the dominant effect, while the moment difference remained more local. The paired damage fields showed the same tendency, with more connected shorter-span-side damage and more visible detached material in the full-span eccentric beams.
\end{enumerate}

These findings show that off-midspan impact should not be evaluated only as a local short-span impact problem. In the more eccentric cases, the full-span beam developed larger shorter-span shear demand and more connected damage than the matched symmetric short-span reference, indicating that the damage may be more severe than what would be expected from the local span length alone. This possibility should be considered in impact assessment of RC beams and examined further for different span ratios, support conditions, reinforcement layouts, impact masses, and material-strength ranges.

\section*{Author Contributions}
Ziqi Gao: Conceptualization, Methodology, Software, Validation, Formal analysis, Investigation, Data curation, Visualization, Writing---original draft preparation; Chi Lu: Conceptualization, Methodology, Software, Writing---review and editing, Supervision, Project administration; Yoshimi Sonoda: Methodology, Supervision, Project administration; Hiroki Tamai: Resources, Supervision.

\section*{Funding}
Not applicable.

\section*{Institutional Review Board Statement}
Not applicable.

\section*{Informed Consent Statement}
Not applicable.

\section*{Data Availability Statement}
Data will be available on request.

\section*{Acknowledgments}
The first author gratefully acknowledges financial support from the China Scholarship Council (CSC).

\section*{Conflicts of Interest}
The authors declare no conflicts of interest.

\section*{Abbreviations}
  The following abbreviations are used in this manuscript:

  \medskip\noindent
  \begin{tabular}{@{}ll}
    CSCM & Continuous surface cap model \\
    FEM & Finite element method \\
    GPU & Graphics processing unit \\
    RC & Reinforced concrete \\
    SPH & Smoothed particle hydrodynamics \\
  \end{tabular}

\bibliographystyle{unsrtnat}
\bibliography{references}

\end{document}